\documentclass[conference]{IEEEtran}
\usepackage{color,graphicx}
\usepackage{url}
\usepackage{amsmath}
\usepackage{algorithmic}
\usepackage{algorithm}
\usepackage{epstopdf}
\usepackage{epsfig}
\usepackage{cite}
\usepackage{mathtools}
\usepackage{breqn}
\usepackage{multirow}
\usepackage{amsfonts}
\usepackage{array}
\usepackage{lipsum}
\definecolor{Gray}{gray}{0.9}
\usepackage[table]{xcolor}

\ifCLASSINFOpdf
\else
\fi

\makeatletter

\newcommand{\Rmnum}[1]{\expandafter\@slowromancap\romannumeral #1@}
\makeatother
\begin{document}
\title{Revisiting 802.11 for User Fairness and Efficient Channel Utilization in Presence of \mbox{LTE-U}}
\author{\IEEEauthorblockN{Anand M. Baswade, Touheed Anwar Atif, Bheemarjuna Reddy Tamma, Antony Franklin A}
\IEEEauthorblockA{Indian Institute of Technology Hyderabad, India\\
Email: [cs14resch11002,ee13b1036,tbr,antony.franklin]@iith.ac.in}}

\maketitle
\begin{abstract}
A promising solution satisfying the industry's demand to have minimum alterations in LTE for its operation in unlicensed spectrum is duty cycled \mbox{LTE-U} scheme, which adopts discontinuous transmission to ensure fair coexistence with 802.11 (\mbox{Wi-Fi}) WLANs.  	
 Even though the scheme guarantees to maintain \mbox{Wi-Fi} network performance, the fairness among \mbox{Wi-Fi} users still remains arcane. In this work, we present a practical scenario where \mbox{LTE-U}, despite being discontinuous (by following an ON/OFF cycle), results in not only unfair throughput distribution among \mbox{Wi-Fi} users but also causes degradation in \mbox{Wi-Fi} AP’s downlink performance. This is due to the domination of few \mbox{Wi-Fi} users who harness channel in both ON and OFF durations of \mbox{LTE-U}, namely non-victim users over those who get access only in OFF duration, called victim users. In this paper, we studied the performance of victim and non-victim \mbox{\mbox{Wi-Fi}} users, and \mbox{\mbox{Wi-Fi}} AP while varying \mbox{LTE-U} ON duration (\emph{i.e.,} duty cycle). A propitious scheme is proposed for WLANs, with regard to ease of implementation, employing Point Coordination Function (PCF) mode of 802.11, promising fairness among \mbox{\mbox{Wi-Fi}} users with improvement in the channel utilization of \mbox{\mbox{Wi-Fi}} network.
 An analytical model is developed to demonstrate guaranteed improvement and validate the simulation results. 
\end{abstract}

\section{Introduction}
An incessant increase in cellular data traffic demand has called upon the traditional telecom operators, to focus on the unlicensed spectrum as a promising solution~\cite{7},\cite{6}. Fair coexistence with other technologies operating in unlicensed spectrum---mainly 802.11 (\mbox{\mbox{Wi-Fi}})---needs to be ensured for successful deployment of \mbox{LTE-U} in different regulatory domains. Though the two schemes, \mbox{LTE-U} Listen Before Talk (LBT) and duty cycled \mbox{LTE-U} are shown to be equally fair with \mbox{Wi-Fi} \cite{10}, the need for quick deployment has forced \mbox{LTE-U} forum~\cite{7} to narrow its focus on markets promoting duty cycled \mbox{LTE-U}. Duty cycled \mbox{LTE-U} is a simple scheme where \mbox{LTE-U} eNB follows duty cycle based discontinuous transmission approach to fairly share an unlicensed channel with \mbox{Wi-Fi} networks operating on the same channel. One such example is Carrier Sense Adaptive Transmission (CSAT) \cite{8}, where \mbox{LTE-U} eNB follows ON and OFF cycle pattern in a given duty cycle period, with ON duration corresponding to \mbox{LTE-U} transmission duration meant to serve \mbox{LTE-U} users in downlink (\mbox{LTE-U} uses unlicensed spectrum only to serve in downlink \cite{7}) and OFF duration corresponding to \mbox{LTE-U} no-transmission duration, to allow \mbox{Wi-Fi} operation. This works well until \mbox{Wi-Fi} AP, the center of communication in an infrastructure Basic Service Set (BSS), is also affected by \mbox{LTE-U} transmissions and hence can serve its users only in \mbox{LTE-U} OFF duration. But in a scenario, where \mbox{Wi-Fi} AP and some users can transmit or receive packets in \mbox{LTE-U} duty cycle (\emph{i.e.,} \mbox{LTE-U} ON period), while remaining users cannot transmit or receive, channel access and consequently the throughput among \mbox{Wi-Fi} users becomes unfair. A similar scenario is considered in \cite{9} where they presented an intercell interference coordination technique when two \mbox{LTE-U} eNBs interfere, but managing a scenario when \mbox{LTE-U} eNB interferes with some users of a \mbox{Wi-Fi} network is unprecedented and is considered as a part of our study in this paper.
 \begin{figure}
 	\includegraphics[totalheight=3.1cm,width=\linewidth]{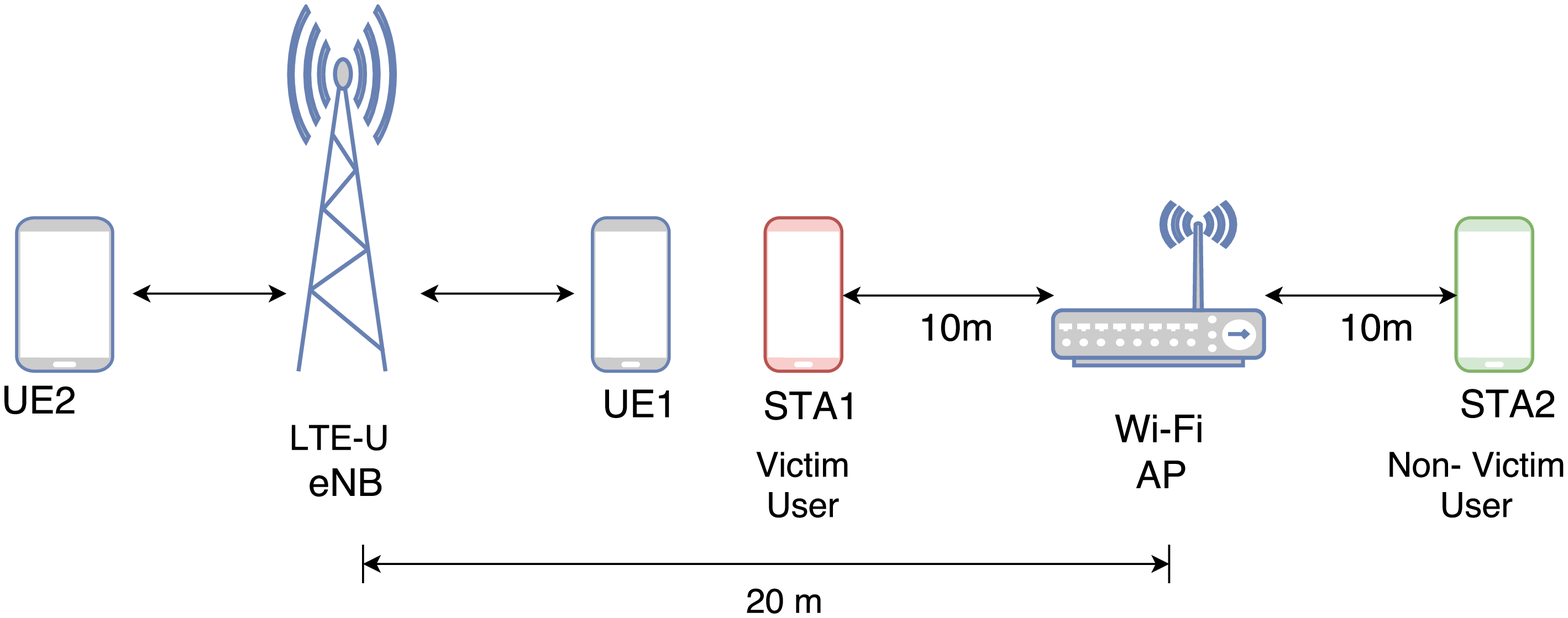}
 	\vspace{-0.4cm}
 	\caption{Motivational example scenario.}\label{1}
 	 \vspace{-0.7cm}
 \end{figure}
 
\begin{figure*}[!htb]
\begin{center}
	\minipage{0.47\textwidth}
	\includegraphics[totalheight=4.2cm,width=\linewidth]{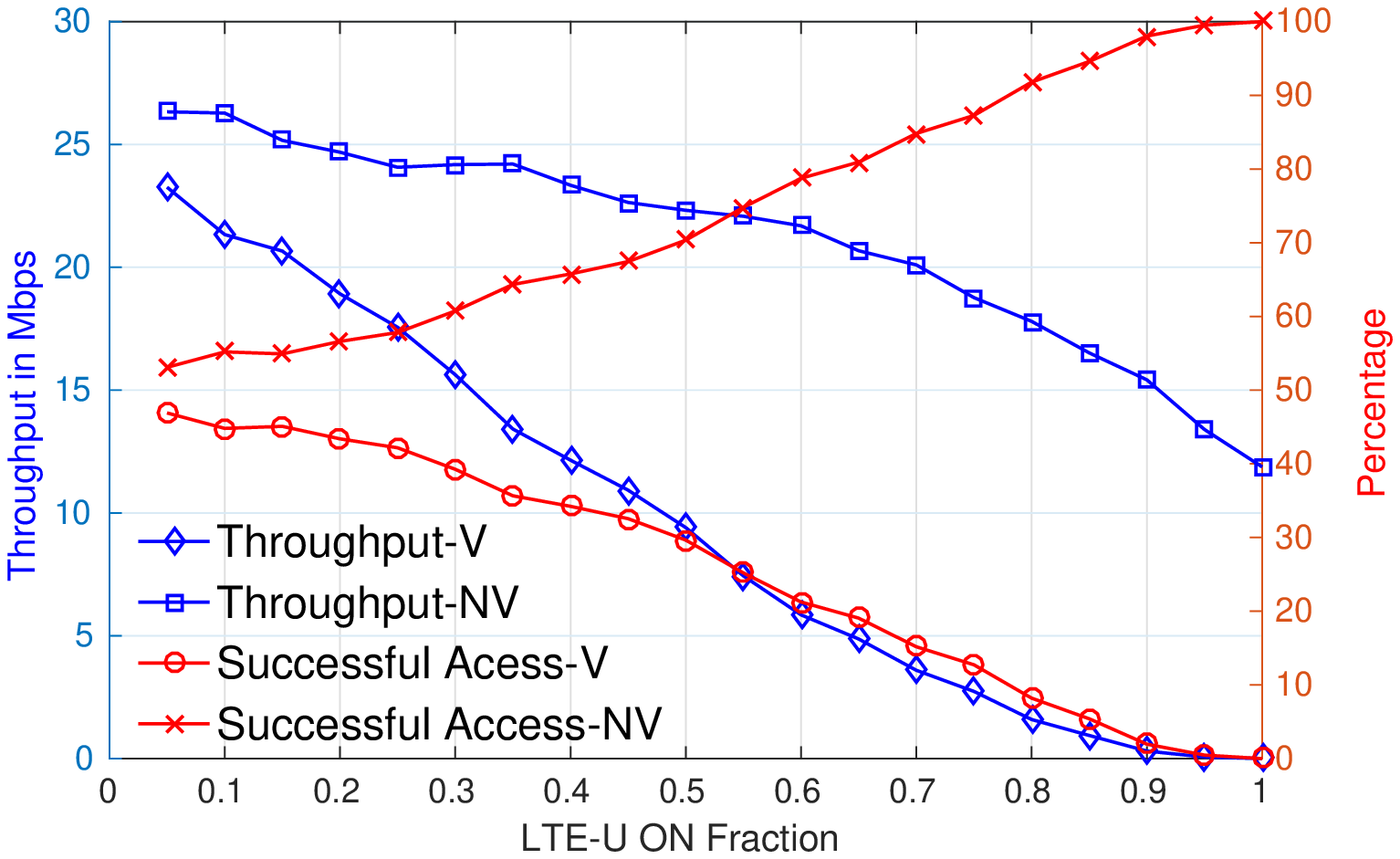}
	\vspace{-0.8cm}
	\caption{\hspace{-0.2cm}Throughput \& successful access (\%) of \mbox{Wi-Fi} user with varying $\eta$.}\label{M1}
	\endminipage\hfill
	~
	\minipage{0.47\textwidth}
	\includegraphics[totalheight=4.2cm,width=\linewidth]{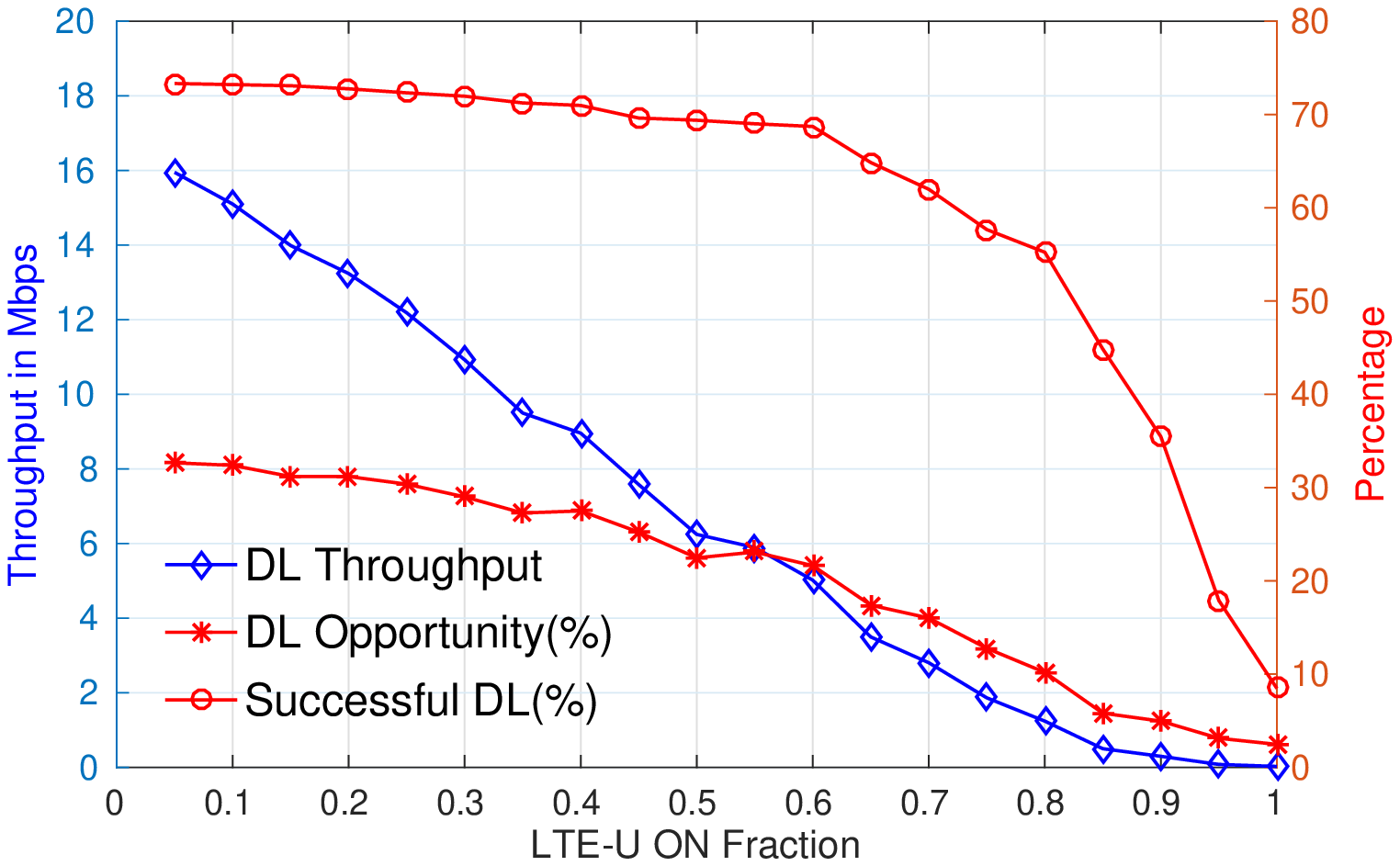}
	\vspace{-0.8cm}
	\caption{\hspace{-0.2cm}\mbox{Wi-Fi} AP performance in DL only in UL \& DL traffic.}\label{M2}
	\endminipage\hfill
\end{center}
\vspace{-0.6cm}
\end{figure*}
 \par Fig. \ref{1} shows a scenario where the effect of \mbox{LTE-U} on \mbox{Wi-Fi} AP is not to an extent that \mbox{Wi-Fi} AP has to defer from channel access during the \mbox{LTE-U} ON period, implying that it can operate in both the \mbox{LTE-U} ON and OFF periods. A \mbox{Wi-Fi} network following Distributed Coordination Function (DCF) mode of operation (referred as standard \mbox{Wi-Fi} for rest of the paper) 
 ensures that every \mbox{Wi-Fi} user has equal chance to access the channel, including the \mbox{Wi-Fi} AP---which is what happens when \mbox{LTE-U} is in OFF period. But when \mbox{LTE-U} is in ON period, \mbox{Wi-Fi} users affected by \mbox{LTE-U}, fail to successfully transmit or receive any data due to two reasons. First, because they sense channel busy due to \mbox{LTE-U} transmissions and second, the interference from \mbox{LTE-U} decreases their Signal to Interference plus Noise Ratio ($ SINR $), below a minimum $ SINR $ required for successful transmission. Note that the successful transmissions in a \mbox{Wi-Fi} network are only the ones which are successfully received and acknowledged, failure to receive a packet or even acknowledge a successfully received packet would eventually be considered as an unsuccessful transmission. We call these users as victim users. On the other hand, remaining users, called as non-victim users, continue to transmit and receive, even during the \mbox{LTE-U} ON period and thus dominate the victim users. In Fig.~\ref{1}, STA1 is a victim user while STA2 is a non-victim user. This domination of non-victim users during the \mbox{LTE-U} ON period causes severe unfairness among the victim and non-victim users. \par Also, every attempt of transmission from \mbox{Wi-Fi} AP to the victim users, during the \mbox{LTE-U} ON period fails and thus elicits multiple retransmissions of the same packet, wasting the channel resources extensively. Hence, our focus in this work, mainly lies in improving the fairness and  channel utilization in the above scenarios by enhancing the throughput of victim users and \mbox{Wi-Fi} AP irrespective of \mbox{LTE-U} duty cycle.
 

\section{Motivational Example}
To study fairness and channel utilization of \mbox{Wi-Fi} in presence of \mbox{LTE-U}, we consider a scenario with an \mbox{LTE-U} eNB and a \mbox{Wi-Fi} AP deployed 20m apart, each with two users as shown in Fig \ref{1}. STA1 and STA2 are two \mbox{Wi-Fi} users with STA1 being a victim user because the interference from \mbox{LTE-U} causes its $ SINR $ to decrease below a minimum required $ SINR $ in \mbox{LTE-U} ON period while STA2 is a non-victim user. We considered a full buffer case for \mbox{Wi-Fi} network, where all \mbox{Wi-Fi} users and the \mbox{Wi-Fi} AP always have data to transmit in UpLink (UL) and DownLink (DL), respectively. For \mbox{LTE-U} network, a similar full buffer case is assumed where unlicensed spectrum is used only for sending DL traffic. \mbox{LTE-U} uses a centralized scheduler while \mbox{Wi-Fi} follows DCF mechanism at MAC layer for sharing radio resources. 

%

If \mbox{LTE-U} detects the presence of a \mbox{Wi-Fi} user (by periodically listening to the channel), \mbox{LTE-U} inherently follows an ON-OFF duty cycle~\cite{7,8}. We varied the fraction of time \mbox{LTE-U} is ON in a duty cycle period, LTE-U ON fraction ($\eta$) from 0 to 1 and observed the performance of \mbox{Wi-Fi} users, namely the victim and non-victim users.
Key observations from our study are: 

\subsubsection{Disproportionate throughput and successful channel access among victim and non-victim users}
 Fig. \ref{M1} shows variation of successful channel access percentage (defined as the ratio of successful attempts of transmitting or receiving data by a user to the total successful attempts by all users in the network) and throughput for each user with $\eta$. It can be observed that the victim user (STA1) has an inevitably less successful access percentage and throughput as compared to the non-victim user (STA2).
 In principle, \mbox{LTE-U} ceases to access the channel, during its OFF cycle, so that the victim users not only get channel access but also transmit and receive their packets satisfying minimum SINR requirement. Non-victim users access to the channel---during this (\mbox{LTE-U} OFF) period---engenders unequal, sometimes very low successful channel access and throughput for the victim users. 

\subsubsection{High retransmission loss because of victim users} Packets from the \mbox{Wi-Fi} AP to victim user (STA1) cannot be successfully received, during the \mbox{LTE-U} ON period, due to their low SINR. This results in retransmission of same packet to the same victim user multiple times, with every retransmission increasing the Contention Window (CW) exponentially, resulting an increase in average Back Off (BO) value of \mbox{Wi-Fi} AP. This increase reduces the channel access opportunity of the \mbox{Wi-Fi} AP. Fig.~\ref{M2} shows a significant decrease in the DL opportunity percentage (defined as the ratio of \mbox{Wi-Fi} AP DL access opportunities to the total access opportunities in \mbox{Wi-Fi} network) and successful DL percentage (defined as the ratio of successful DL transmissions to the total DL transmissions by AP) with increase in $\eta$.
Fig. \ref{M2} also shows a decrease in DL throughput of \mbox{Wi-Fi} AP, which is a result of decreased DL access opportunities and successful DL transmissions.    
\section{System Model}\label{S3}
We consider a scenario where \mbox{LTE-U} eNB and \mbox{Wi-Fi} AP from same or different operators coexist on the same unlicensed
channel with the same amount of bandwidth. The users are associated with either \mbox{LTE-U} or \mbox{Wi-Fi}, with \mbox{LTE-U} following
a duty cycle based scheme for fair coexistence with \mbox{Wi-Fi}. The fraction of time \mbox{LTE-U} uses for transmission $ (\eta) $ in duty
cycle period can be fixed or adaptively adjusted depending on \mbox{Wi-Fi} channel utilization \cite{7,8}. To make proper and efficient use of unlicensed spectrum, it is important to have some kind of (indirect) coordination between \mbox{LTE-U} and \mbox{Wi-Fi} AP devices. Hence, we assume that there is an inter-RAT controller as in \cite{3,1}. The inter-RAT controller can be implemented by the operator if both \mbox{LTE-U} and \mbox{Wi-Fi} AP are deployed by the same operator or it can be implemented by a third party vendor to extend the service among different operators. The controller communicates with \mbox{LTE-U} eNB and \mbox{Wi-Fi} AP
to synchronize the \mbox{LTE-U} duty cycle period with the \mbox{Wi-Fi} beacon interval (in multi-eNBs and multi-APs scenario, the controller will synchronize all the \mbox{Wi-Fi} APs to a single beacon interval, and all the \mbox{LTE-U} eNBs also to the same beacon interval).  Beacons in \mbox{Wi-Fi} are the management frames, sent by the \mbox{Wi-Fi} AP periodically, containing the network information, and beacon interval is the frequency at which these beacons are sent. After synchronization, if AP changes the beacon interval, it will re-advertise the beacon transmission time and the beacon interval to LTE-eNB via the controller. Now it is made as the responsibility of eNB to match its duty cycle period to the beacon interval (i.e., LTE OFF time + LTE ON time = beacon interval) and always start the OFF period at the beginning of new beacon interval. \mbox{LTE-U} eNB also informs the fraction of time it is ON ($\eta$) to the AP via the controller and allows \mbox{Wi-Fi} AP to improve channel utilization. 
\begin{figure}[htb!]
\begin{center}
\includegraphics[totalheight=3.1cm,width=8.0cm]{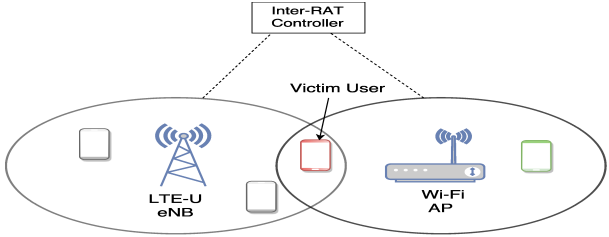}
\caption{System model.}
\label{MSD}
\end{center}
\vspace{-0.5cm}
\end{figure}
Fig. \ref{MSD} shows the considered system model where \mbox{LTE-U} cell and \mbox{Wi-Fi} BSS are partially overlapped. Workflow of the system model is as follows: \mbox{Wi-Fi} user informs to the AP whenever it detects any non \mbox{Wi-Fi} signal (with its capability to differentiate \mbox{Wi-Fi} and non-\mbox{Wi-Fi} signals using \mbox{Wi-Fi} preamble) causing interference to an extent that its minimum SINR requirement is not satisfied to establish communication. Whenever \mbox{Wi-Fi} AP has a victim user, the AP reports to the inter-RAT controller and starts following a co-ordination approach. Cellular users are not considered as victim users because LTE eNB can serve them using the licensed spectrum whereas \mbox{Wi-Fi} AP has only unlicensed spectrum to serve its users.
\section{Proposed Work}
Observations from the motivational example persuades one, for a need to adopt a scheme to ensure the fairness among \mbox{\mbox{Wi-Fi}} users. The DCF mechanism during both the \mbox{LTE-U} ON and OFF durations, without any intelligent steps to improve the network performance---despite having throughput knowledge of the users---can be identified as the main bottleneck. Our work identifies this bottleneck, and then proposes an intelligent and controlled use of Point Coordination Function (PCF) mechanism, without much modification to the existing system, to address this issue and thereby harnesses maximum~gain.
\subsection{Use of Point Coordination Function}
\begin{figure}[htb!]
\begin{center}
\includegraphics[totalheight=3.1cm,width=8.8cm]{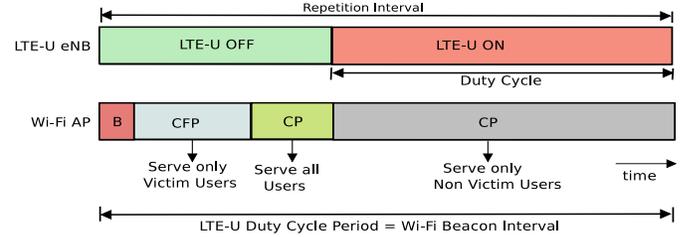}
\caption{Use of PCF mode for victim users.}
\label{PCF}
\end{center}
\vspace{-0.5cm}
\end{figure}

The scenario presented in the above system model, requires a control in the channel access of non-victim users accompanied by a special channel access given to the victim users. To accomplish the above goal, we advocate usage of PCF mode of \mbox{Wi-Fi}, where the channel is divided into two periods---Contention Free Period (CFP) and Contention Period (CP)---with the difference being that the CP involves contention among users whereas CFP requires only those users to be served which are polled or selected by the \mbox{Wi-Fi} AP. Since, victim users cannot be served in the \mbox{LTE-U} ON period, our proposed scheme allows the \mbox{Wi-Fi} AP to transmit and receive data from victim users during the \mbox{LTE-U} OFF period---by using CFP. After the victim users are served, for the remaining \mbox{LTE-U} OFF period, AP follows CP (\emph{i.e., DCF}) to serve all the users, with each user contending and gaining access to the medium. With the onset of \mbox{LTE-U} ON period, \mbox{Wi-Fi} network still continues to be in CP, but \mbox{Wi-Fi} AP having the additional knowledge of victim users, does not attempt to serve the victim users. This prevents unnecessary re-transmissions and helps the \mbox{Wi-Fi} AP, maintain its channel access opportunity same as non-victim users. The complete operation is illustrated in Fig.~\ref{PCF}. Also, we propose that the \mbox{Wi-Fi} AP sends its beacons at the beginning of \mbox{LTE-U} OFF period. This ensures that every \mbox{Wi-Fi} user listens to them, which is important for a user to maintain its association with the \mbox{Wi-Fi} network.

\subsection{Optimum value of CFP}
If the entire \mbox{LTE-U} OFF period was used to serve the victim users (by making entire \mbox{LTE-U} OFF duration as CFP), the result can be, victim users gaining undue advantage in terms of throughput. To avoid this, we find the minimum CFP duration, within the \mbox{LTE-U} OFF period, that is required for ensuring throughput fairness among the \mbox{Wi-Fi} users. If $x$ is the duration to operate in CFP, variation in $x$ can give control of how much the victim users has perquisite over the non-victim. In the case when $x$ is made as zero, the proposed scheme degenerates to standard \mbox{Wi-Fi}. Any value of $x$ between $0$ to $\eta$ will give better throughput for the victim users (if present) as well for the entire \mbox{Wi-Fi} setup. One approach to realize optimum CFP duration is to adjust $ x $ by observing the throughputs of previous duty cycle period. The updated CFP duration $ (T_{cfp}) $ in terms of previous CFP duration $ T_{cfp}^{old} $ can be given as
\begin{equation}
T_{cfp} = \frac{\Gamma^{new}_{nv}}{\Gamma^{new}_{v}}*T^{old}_{cfp}
\end{equation}
$\Gamma^{new}_{nv}$ and $\Gamma^{new}_{v}$ are calculated using average throughputs of non-victim and victim users, respectively as
\begin{eqnarray}
\Gamma^{new}_{nv} = (1- \alpha)*\Gamma^{prev}_{nv}+ \alpha * \Gamma^{old}_{nv} \\
\Gamma^{new}_{v} = (1 - \alpha)*\Gamma^{prev}_{v}+ \alpha*\Gamma^{old}_{v} 
\end{eqnarray}
Where $\Gamma^{prev}_{nv}$ and $\Gamma^{prev}_{v}$ are average throughputs in previous duty cycle period and $\Gamma^{old}_{nv}$ and $\Gamma^{old}_{v}$ are the $\Gamma^{new}_{nv}$ and $\Gamma^{new}_{v}$ of the previous duty cycle period, for non-victim and victim user, respectively and $\alpha$ is a smoothing parameter.

\section{Throughput Analysis of Proposed Scheme}
For analyzing throughput of proposed scheme, we consider an \mbox{LTE-U} eNB and a \mbox{Wi-Fi} AP, each having $N_t$ clients. \mbox{Wi-Fi} throughput with PCF mode enabled (proposed scheme) differs extensively from its standard mode of operation (\emph{i.e.,} a \mbox{Wi-Fi} network with nodes configured in DCF mode). 
We compare the throughput of proposed scheme with that of the standard \mbox{Wi-Fi} operation. \mbox{LTE-U} is assumed to serve its unlicensed users in DL \cite{6}, and \mbox{Wi-Fi} is studied for the following two scenarios:
\begin{enumerate}
	\item DL only traffic from the AP to its users.
	\item UL and DL traffic at all nodes in the network.
\end{enumerate}
\subsection{DL only traffic with full buffer }
Let $T_{difs}$, $T_{slot}$, and $T_{sifs}$ denote the DIFS interval, back-off slot time, and SIFS interval for the \mbox{Wi-Fi} network, respectively. Let $T_{busy}$ denote the amount of time channel will be busy after a data transmission start. Therefore, $T_{busy} = T_{rts} + T_{sifs} + T_{cts} + T_{sifs} + T_{data} + T_{sifs} + T_{ack}$, where $T_{data}$ represents the average time required to successfully transmit a data packet - when \mbox{LTE-U} is ON and $T_{ack}$ represents the average time required to transmit an ACK. 
Let $Q_K$ denote the random back-off value selected by the \mbox{Wi-Fi} AP during the $K^{th}$ transmission and $T_{succ}$ denotes the average  time taken by the AP to transmit one data burst successfully. Since a DL only traffic will not elicit any collisions, $P(Q_K=q_k) = \frac{1}{W}$, where W denotes the minimum back-off window size and
average value of $q_k$ is given as $ E[q_k] = \cfrac{W-1}{2} $ then $T_{succ}$ is given by
\begin{eqnarray}\label{T_succ}
T_{succ} &=&  E[T_{difs} + T_{slot}*q_k + T_{busy}] \nonumber \\
         &=& T_{difs} + \cfrac{W-1}{2}*T_{slot} + T_{busy}
\end{eqnarray}

To determine the throughput of such scenario, we employ slot based approach, which assumes that \mbox{Wi-Fi} AP follows a process of transmitting the data on a per slot basis. A slot in essence, is determined by the time it takes to transmit to a victim or a non-victim user.

During the LTE-ON period, successful transmission to a victim user is indeed not possible (due to low SINR). So the \mbox{Wi-Fi} AP will keep on sending Request-to-Send (RTS) packets, until the number of attempts reaches the Short Retry Limit $(R_{limit})$, before it discards the packet.  Nonetheless, a non-victim user can be served successfully during this period (assuming an ideal channel). Therefore, on an average a successful transmission will last for $T_{succ}$ and an unsuccessful, involving Clear-to-Send (CTS) time out duration ($T_{cts\_timeout}$), $R_{limit}$ number of re-transmissions and exponential growth of CW with every re-transmission,  
will last for $T_{un-succ}$, where  
\begin{equation}\label{T_unsucc}
\begin{split}
T_{un-succ} =  \cfrac{1}{2}\Big[2^{m}W\Big(R_{limit}-m+1\Big) - R_{limit}-W\Big]\\ 
+ R_{limit} (T_{difs} + T_{rts} + T_{cts\_timeOut})
\end{split}
\end{equation}
and, $m$ denotes $ \log_{2}{\cfrac{CW_{max}}{W}}$. 
Let $T_{i}$ be a random variable representing the time consumed by any slot, given by
\begin{equation}
T_i = \left \{
\begin{tabular}{ccc}
$T_{un-succ}$ & \text{with prob. } $p$ \\
$T_{succ}$ & \text{with prob. } $1-p$ 
\end{tabular}
\right \}
\end{equation} 
\begin{equation}
\text{and} \quad E[T_i] = p*T_{un-succ} + (1-p)*T_{succ}
\end{equation}
where $ p = \cfrac{N_v}{N_t}$, assuming that AP randomly selects a user to serve, with no special preferences.\\
The total time taken by $n$ such slots will be given by \[T = \sum_{i=1}^{n}{T_i} \implies E[T] = E[\sum_{i=1}^{n}{T_i}] = \sum_{i=1}^{n}{E[T_i]}\] 
Since, $T_{i}'s$ are Independent and Identically Distributed (IID) copies of each other, 
\begin{equation}
 E[T] = n*E[T_i] 
\end{equation}
Using $T_{succ}$ and $T_{un-succ}$, we categorize the performance of standard \mbox{Wi-Fi} and proposed scheme to show the benefits in terms of fairness achieved among \mbox{Wi-Fi} users and throughput improvement in \mbox{Wi-Fi} network. 
\subsubsection{Standard \mbox{Wi-Fi}} 
Since the behavior of \mbox{Wi-Fi} during the LTE-ON and LTE-OFF periods is quite dissimilar, we analyze both of them separately. During the LTE-OFF period, all the \mbox{Wi-Fi} transmissions are successful, implying that there is no randomness involved in the slot duration ($T_{i}$). \[T_i = T^'_{succ} \quad \text{and} \quad E[T] = n*T^'_{succ}  \]
$ T_{succ}^' $ differs from $ T_{succ} $ in $ T_{data} $; no interference from \mbox{LTE-U} causes \mbox{Wi-Fi} AP to use higher Modulation and Coding Scheme (MCS) and thus will decrease $ T_{data} $, and $ T_{succ}^' < T_{succ}$.  If average size of packet is $E[P]$ (bytes), then the total number of bytes transmitted during this interval will be $n*E[P]$ and the throughput of network during the LTE-OFF period $(\Gamma_{off}^S)$ will be
\begin{equation}
\Gamma_{off}^S = \cfrac{n*E[P]}{n*T^'_{succ}} = \cfrac{E[P]}{T^'_{succ}}
\end{equation} 
The victim user $(\Gamma_{off,v}^S)$ and non-victim user $(\Gamma_{off,nv}^S)$ throughputs will be same and are given by
\begin{equation}
\Gamma_{off,v}^S = \Gamma_{off,nv}^S = \cfrac{\Gamma_{off}^S}{N_t} = \cfrac{1}{N_t}*\cfrac{E[P]}{T^'_{succ}}
\end{equation}
During the LTE-ON period, AP will be able to successfully transmit to a non-victim user, but its transmission to a victim user, will be unsuccessful. This will cause $T_i$ to be a random variable, fluctuating between $T_{succ}$ and $T_{un-succ}$, given by Eqn~(\ref{T_succ}) and Eqn~(\ref{T_unsucc}).
The average duration to complete $n$ transmissions ($E[T]$) during this interval is given by
\begin{equation}
E[T] = nE[T_i] = n\Big(pT_{un-succ} + (1-p)T_{succ}\Big)
\end{equation}
Data packet is transmitted only when a successful transmission takes place, \emph{i.e.,} if the packet is meant for a non-victim user, implying that the average amount of data successfully transmitted is the sum of all transmissions to the non-victim users in the \mbox{Wi-Fi} network. Therefore, the average amount of data transmitted is $n(1-p)E[P]$.
The total throughput $\Gamma_{on}^S$ can then be given by 
\begin{equation}
 \Gamma_{on}^S = \cfrac{(1-p)E[P]}{\Big(pT_{un-succ} + (1-p)T_{succ}\Big)}
\end{equation}
Since the victim users do not receive any data, their throughputs  $(\Gamma_{on,v}^S)$ will be zero $(\Gamma_{on,v}^S = 0)$,
while the non-victim users will equally share the total throughput as 
\begin{equation}\label{19}
 \Gamma_{on,nv}^S = \cfrac{1}{N_{t} - N_v}\cfrac{(1-p)E[P]}{\Big(pT_{un-succ} + (1-p)T_{succ}\Big)}
\end{equation}
The total throughput of a victim user ($\Gamma_v^S$) and that of a non-victim user ($\Gamma_{nv}^S$) in a duty cycle period  are given by
\begin{equation}\label{th_V_S}
\Gamma_v^S = (\eta) \Gamma_{on,v}^S + (1-\eta)\Gamma_{off,v}^S
       = (1-\eta)\cfrac{1}{N_t}*\cfrac{E[P]}{T_{succ}^'} 
\end{equation}
\begin{equation}
\text{Similarly, }\Gamma_{nv}^S =(\eta)\Gamma_{on,nv}^S + (1-\eta)\cfrac{1}{N_t}*\cfrac{E[P]}{T_{succ}^'}
\end{equation}
Hence, the total \mbox{Wi-Fi} throughput ($\Gamma^{S}$) of standard \mbox{Wi-Fi} network in the presence of \mbox{LTE-U} is 
\begin{eqnarray}\label{th_S}
\Gamma^{S}\hspace{-0.2cm} &=&\hspace{-0.2cm} (N_v)\Gamma_v^{S} + (N_t-N_v)\Gamma_{nv}^{S} \nonumber \\
       &=&\hspace{-0.2cm} (\eta)(N_t-N_v)(\Gamma_{on,nv}^S) + \Big((1-\eta)\cfrac{E[P]}{T_{succ}^'}\Big) 
\end{eqnarray}

\subsubsection{Proposed scheme}
Using the information, regarding the presence of victim users and \mbox{LTE-U} ON period, in our proposed scheme (for efficient utilization of unlicensed spectrum) the \mbox{Wi-Fi} AP intelligently defers from transmitting to the victim users during \mbox{LTE-U} ON period, and serves these users using PCF mode during \mbox{LTE-U} OFF period.
Since LTE-ON period involves only the non-victim users transmissions, every transmission will be successful and will last for $ T_{succ} $ duration. Hence, the throughput of each non-victim ($ \Gamma_{on,nv}^{PS} $) and victim $ (\Gamma_{on,v}^{PS}) $ user is given by 
\begin{equation}
 \Gamma_{on,v}^{PS} = 0 \quad \text{and} \quad \Gamma_{on,nv}^{PS} = \cfrac{1}{N_{t}-N_v}*\cfrac{E[P]}{T_{succ}}
\end{equation}
But, during the LTE-OFF period, the behavior of proposed scheme is quite ingenious. Let $T_{cfp}$  denote the average time taken to serve a victim user, in LTE-OFF period using CFP. The time taken to transmit a data packet (and receive an ACK) using CFP is given by
\begin{equation}
T_{cfp} =  T_{sifs} + T_{data} + T_{sifs} + T_{ack}
\end{equation}
Assuming average size of packet as $E[P]$, time taken for $n$ such transmissions ($T^n_{cfp}$) and throughput of total network ($\Gamma_{off}^{cfp}$) are given by
\begin{equation}
T^n_{cfp} = \sum_{i=1}^{n}{T_{cfp}} = n*T_{cfp} \quad \text{and} \quad \Gamma_{off}^{cfp} = \cfrac{E[P]}{T_{cfp}}
\end{equation}
As none of the non-victim users are being served in the CFP, their throughput ($\Gamma_{off,nv}^{cfp}$) will be zero, while the victim users ($\Gamma_{off,v}^{cfp}$) equally share the total throughput as 
\begin{equation}\label{26}
\Gamma_{off,nv}^{cfp} = 0 \quad \text{and} \quad \Gamma_{off,v}^{cfp} = \cfrac{1}{N_v}*\cfrac{E[P]}{T^{cfp}}
\end{equation}
With the remaining time in which the LTE is still OFF, \mbox{Wi-Fi} uses PCF mode in CP to serve all the users, which will have the throughput, same as that of standard \mbox{Wi-Fi} operating in LTE-OFF period ($\Gamma_{off}^S$). Hence, the \mbox{Wi-Fi} throughput during the LTE-OFF period within CP ($\Gamma_{off}^{cp}$) is given as
\begin{figure*}[!htb]
\minipage{0.48\textwidth}
\includegraphics[totalheight=4.4cm,width=\linewidth]{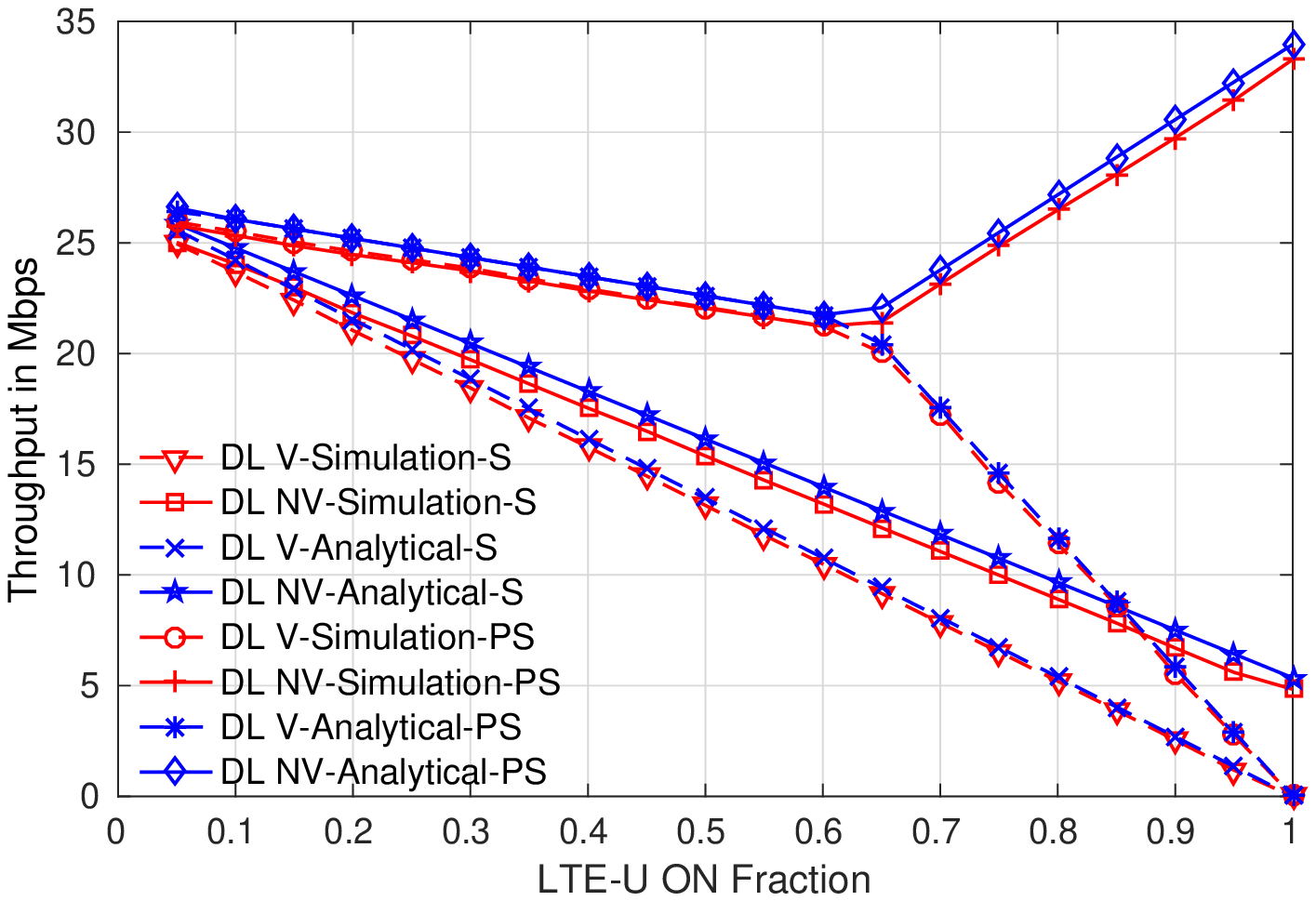}
\vspace{-0.8cm}
\caption{\hspace{-0.2cm}Throughput of both the schemes in DL only traffic with varying $\eta$.}\label{R3}
\endminipage\hfill
~
\minipage{0.48\textwidth}
  \includegraphics[totalheight=4.4cm,width=\linewidth]{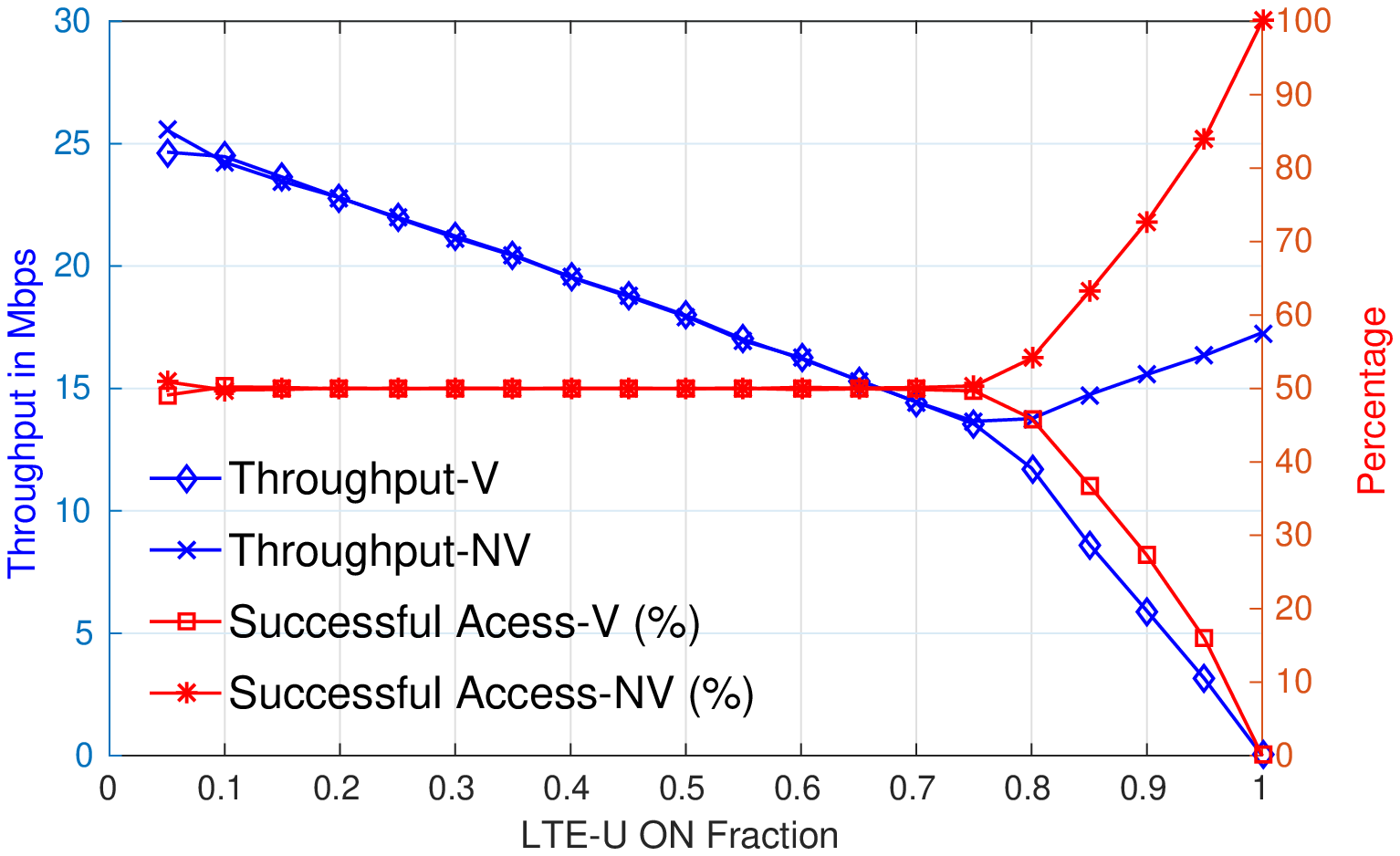}
  \vspace{-0.8cm}
  \caption{\hspace{-0.2cm}\ Throughput and access(\%) of proposed scheme in UL \& DL traffic.}\label{R1}
\endminipage\hfill
\vspace{-0.6cm}
\end{figure*}

\begin{equation}
\Gamma_{off}^{cp} = \Gamma_{off}^S
\end{equation}
Also, each victim ($\Gamma_{off,v}^{cp}$) and non-victim user ($\Gamma_{off,nv}^{cp}$) will have equal throughputs as
\begin{equation}
\Gamma_{off,v}^{cp} = \Gamma_{off,nv}^{cp} = \cfrac{ \Gamma_{off}^S}{N_t}
\end{equation}
Assuming that the fraction of time, \mbox{Wi-Fi} operates in CP, during the \mbox{LTE-U} OFF period, as $x \leq (1-\eta)$, the total average throughput of each victim ($\Gamma_v^{PS}$) and non-victim user ($\Gamma_{nv}^{PS}$), with the proposed scheme can be given as
\begin{eqnarray}\label{th_NV_PS}
\Gamma_v^{PS} \hspace{-0.2cm} &=&\hspace{-0.2cm} (\eta)\Gamma_{on,v}^{PS}  + (1- \eta - x) \Gamma_{off,v}^{cp} + (x) \Gamma_{off,v}^{cfp} \nonumber\\
         &=&\hspace{-0.2cm} (1- \eta - x)\cfrac{1}{N_t}\cfrac{E[P]}{T_{succ}^'} + (x)\cfrac{1}{N_v}\cfrac{E[P]}{T_{cfp}}
\end{eqnarray}
Similarly,
\begin{eqnarray}
\Gamma_{nv}^{PS}\hspace{-0.2cm} &=&\hspace{-0.2cm} (\eta)\Gamma_{on,nv}^{PS} + (1 - \eta - x) \Gamma_{off,nv}^{cp} + (x) \Gamma_{off,nv}^{cfp} \nonumber\\
             &=&\hspace{-0.2cm} (\eta)\cfrac{1}{N_{t}-N_v}\cfrac{E[P]}{T_{succ}} + (1-\eta-x)\cfrac{ 1}{N_t}\cfrac{E[P]}{T_{succ}^'}
\end{eqnarray}
Hence, the total \mbox{Wi-Fi} throughput $(\Gamma^{PS})$ in our proposed scheme is given by 
\begin{eqnarray}\label{th_PS}
\Gamma^{PS} \hspace{-0.2cm} &=& \hspace{-0.2cm} (N_v)\Gamma_v^{PS} + (N_t-N_v)\Gamma_{nv}^{PS} \nonumber \\ 
        &=& \hspace{-0.2cm}(\eta)\cfrac{E[P]}{T_{succ}} + (1-\eta - x)\cfrac{E[P]}{T_{succ}^'} + (x)\cfrac{E[P]}{T_{cfp}}
\end{eqnarray}

%
%

\subsubsection{Observations and gains}
    $ (i) $ \textit{Throughput of each victim user will increase irrespective of $ \eta $ :} Using Eqn~(\ref{th_V_S}) and Eqn~(\ref{th_NV_PS}), $ \Gamma_v^{PS} $ can be re-written as
    \begin{equation}\label{O1}
    \Gamma_v^{PS} = \Gamma_v^{S} + (x)\Big(\cfrac{E[P]}{N_v * T_{cfp}} - \cfrac{E[P]}{N_t * T_{succ}^'}\Big)
    \end{equation}
    The second term on the Right Hand Side (RHS) of Eqn~(\ref{O1}) indicates whether the proposed scheme can outperform standard \mbox{Wi-Fi} in terms of throughput of victim user. The two fractions within the second term are the throughputs of victim user in the $ x $ duration with the proposed scheme and standard \mbox{Wi-Fi}, respectively.\\
    Since, $T_{cfp} < T_{succ}^'$ and $N_v \leq N_t$,
    
    \begin{equation}
    \cfrac{1}{N_v}\cfrac{E[P]}{T_{cfp}} > \cfrac{1}{N_t}\cfrac{E[P]}{T_{succ}^'} \implies \Gamma_v^{PS} > \Gamma_{v}^{S}
    \end{equation}

    $ (ii) $ \textit{Total throughput of \mbox{Wi-Fi} network will increase regardless of $ \eta $:} Using Eqns~(\ref{th_S}) and (\ref{th_PS}), $ \Gamma^{PS} $ can be written as
    \begin{eqnarray}
        \Gamma^{PS}\hspace{-0.2cm} &=&\hspace{-0.2cm} \Gamma^{S} + x\Big(\cfrac{E[P]}{T_{cfp}} - \cfrac{E[P]}{T_{succ}^'}\Big) +\\
                & &\hspace{-0.2cm} \eta\Big(\cfrac{E[P]}{T_{succ}} - \cfrac{E[P]}{\Big(\cfrac{p}{1-p}T_{un-succ} + T_{succ}\Big)} \Big) \nonumber     
    \end{eqnarray}
    Since, $ T_{un-succ} > 0 $ and $ T_{succ}^' > T_{cfp} $, implies that all the terms on the RHS are positive. Therefore, $ \Gamma^{PS} > \Gamma^{S} $.
\subsection{UL and DL full buffer scenario}
Here, we assume that all the \mbox{Wi-Fi} nodes including the AP are saturated. 
Achieving fairness among \mbox{Wi-Fi} users becomes much more challenging---with non-victim users transmitting huge amount of data in UL.
During the LTE-ON period, the \mbox{Wi-Fi} AP as well as the victim users will increase their CWs, due to mutual packet loss, whereas, the non-victim users will maintain a reasonable CW size and hence, access the medium very frequently. Since most of these accesses are successful (except for those which collide), their UL throughput will drastically increase, resulting in further unfairness. The network throughput might seem to be reasonable, but the DL throughput can reach a very low value 
(due to the increase in \mbox{\mbox{Wi-Fi}} AP's CW), which is unpalatable in many scenarios. Using the proposed scheme, \mbox{Wi-Fi} AP defers from transmitting to victim users during \mbox{LTE-U} ON period, implying that its CW remains similar to that of non-victim users and hence, UL and DL throughputs remain evenly distributed.

During the \mbox{LTE-U} OFF period, the proposed scheme would increase the throughput of victim users, similar to the DL only scenario, with \mbox{Wi-Fi} AP sending and receiving data from victim users in CFP.
 An optimum value of the CFP duration will be reached and in each duty cycle period, the throughputs of victim and non-victim users would become identical. Further, a good uniformity between the UL and DL traffic would be achieved in the network.  
\begin{figure*}[!htb]
\begin{center}
\minipage{0.47\textwidth}
  \includegraphics[totalheight=4.4cm,width=\linewidth]{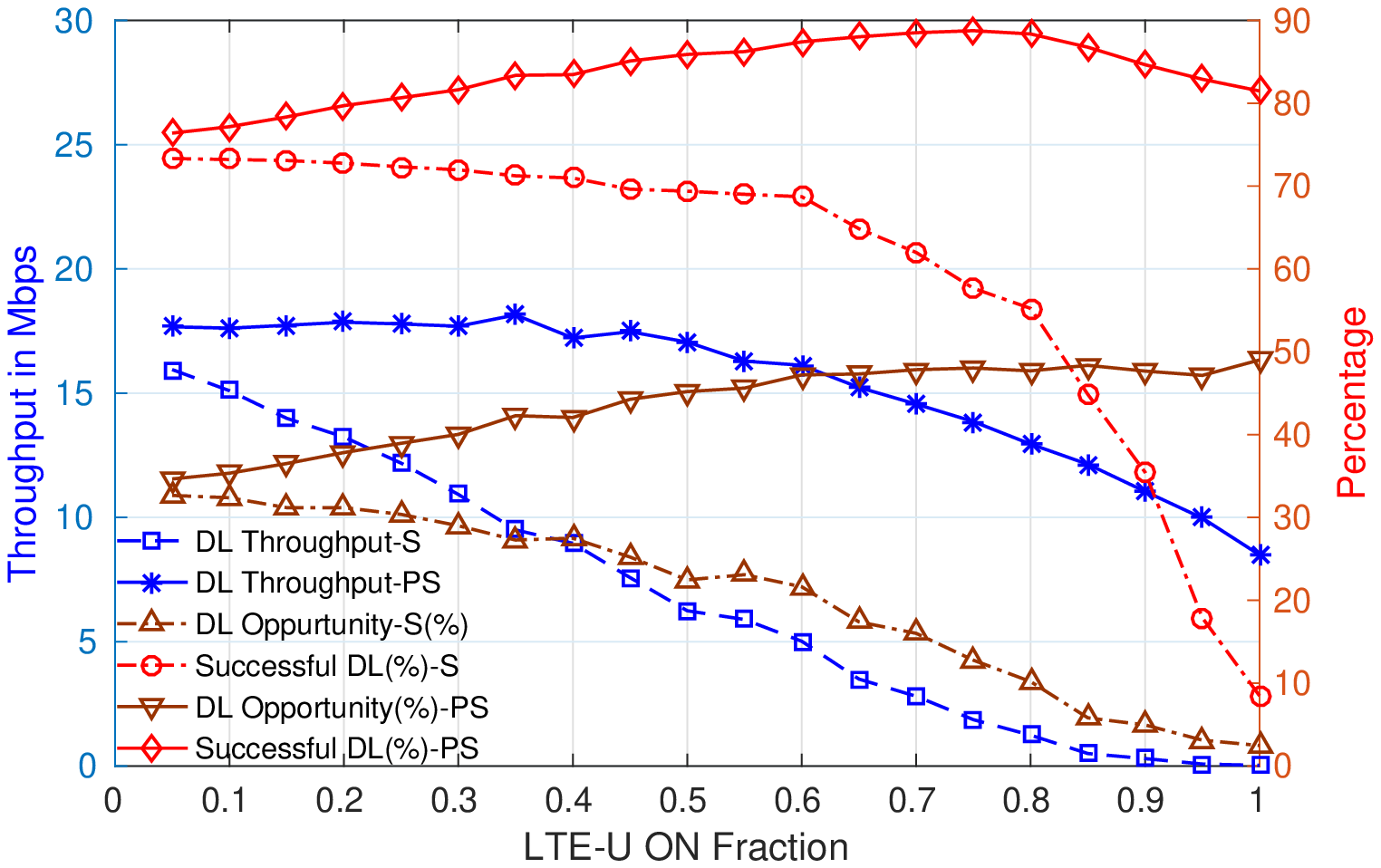}
  \vspace{-0.8cm}
  \caption{\hspace{-0.2cm}\mbox{Wi-Fi} AP performance in DL only with UL \& DL traffic.}\label{R2}
\endminipage\hfill
~
\minipage{0.47\textwidth}
  \includegraphics[totalheight=4.4cm,width=\linewidth]{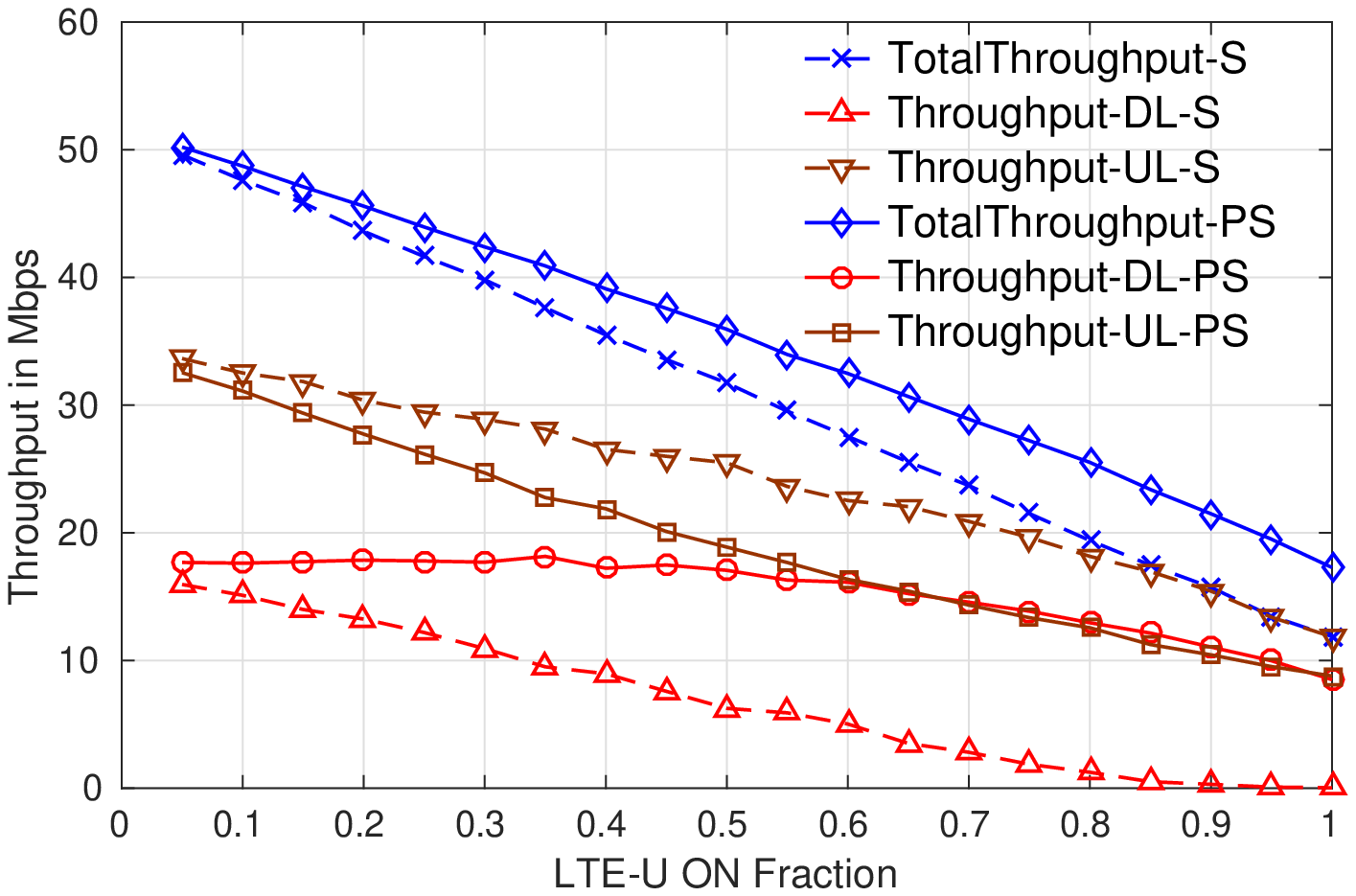}
  \vspace{-0.8cm}
  \caption{\hspace{-0.2cm}Throughput of both the schemes in UL \& DL traffic with varying~$\eta$.}\label{R1_2}
\endminipage\hfill
\vspace{-0.6cm}
\end{center}
\end{figure*} 

\section{Performance Evaluation}
The system model described in Section~\ref{S3} has been simulated using MATLAB with parameters given in Table \ref{Parameter1}. For simulation, we considered the same scenario as shown in Fig.~\ref{1}. Results with standard \mbox{Wi-Fi} are compared with the proposed scheme by simulating for 10 seconds and averaging over 10 different seed values. As the focus is to improve performance of victim users and \mbox{Wi-Fi} network, irrespective of duty cycle, the results are shown for the same.
\begin{table}[h]
\vspace{+0.2cm}
\caption{\mbox{Wi-Fi} \& \mbox{LTE-U} Parameters}
\vspace{-0.2cm}
\centering
\begin{tabular}{|c|c|c|c|}
\hline
\multicolumn{2}{|c}{\bfseries \mbox{Wi-Fi} parameters}&\multicolumn{2}{|c|}{\bfseries Common Parameters}\\
\hline
\bfseries Parameters & \bfseries Values & \bfseries Parameter & \bfseries Value \\
\hline
$CW_{min}$,$CW_{max}$ & 16,1024 & Tx Power & 20 dBm \\
\hline
PHY, MAC Header & 128,272 bits &Operating Freq. & 5.3 GHz \\
\hline
ACK,RTS & 240,288 bits & Noise & -101 dBm \\
\hline
Payload,MPDU & 8148 bits,4 & Bandwidth & 20 MHz \\
\hline
Slottime,CTStimeout & 9,50 $\mu$s & Antenna Ht. & 10 meter \\
\hline
DIFS, SIFS& 34,16 $\mu$s & User Antenna Ht. & 1 meter\\
\hline
Beacon Interval,$\alpha$ & 100 ms,0.5 & Antenna& SISO\\
\hline
\end{tabular}
\begin{tabular}{|p{3.09cm}| p{4.8cm}|}
\multicolumn{1}{|c}{\bfseries Parameter} & \multicolumn{1}{|c|}{\bfseries Value} \\
\hline
\mbox{Wi-Fi} PHY Rates (Mbps)& 6.5 13, 26, 39, 52, 78, 104, 117, 130  \\[.2ex]
\hline
Required SNR (in dB) &2, 5, 7, 9, 13, 17, 20, 22, 23\\[.2ex]
\hline

Traffic & Full buffer via saturated UDP flows \\
\hline
Channel & No shadow/Rayleigh fading \\
\hline
Path Loss Model \cite{PL} & 36.7log10(d[m])+22.7+26log10(freq[GHz])  \\
\hline
\end{tabular}
\label{Parameter1}
\vspace{-0.4cm}
\end{table}

%
\vspace{-0.1cm}
\subsection{DL traffic with full buffer}
Fig. \ref{R3} shows the throughputs of victim and non-victim user with the proposed scheme (PS) and Standard \mbox{Wi-Fi} (S) in DL only scenario. The figure has both simulation and analytical plots for varying $ \eta $. It shows that the throughputs of victim and non-victim user are not just higher (compared to standard \mbox{Wi-Fi}) but also are identical, thereby achieving fairness among \mbox{Wi-Fi} users, all the way from $ \eta=0 $ (\mbox{LTE-U} completely OFF) to $ \eta < \eta_t $. We define $ \eta_t $ as the maximum value of $ \eta $ for which throughput fairness is achievable.
For $ \eta > \eta_t $, the \mbox{LTE-U} OFF duration (\emph{i.e.,} $ 1-\eta $) is insufficient to provide the victim user with same throughput as that of the non-victim user, in spite of configuring the complete \mbox{LTE-U} OFF period as CFP. 
 
 \par Improvement in victim users throughput can be attributed to the use of CFP by the \mbox{Wi-Fi} AP to serve them, and improvement in non-victim users throughput can be imputed to the fact that \mbox{Wi-Fi} AP does not waste the channel in transmitting to the victim users during the LTE-ON duration and serves only the non-victim users. This further helps the network to maintain throughput, especially in scenarios where $ \eta > \eta_t $.
From both simulation and analytical results, it is clearly conveyed that our proposed scheme performs better in achieving fairness for $ \eta \leq \eta_t $, and improving channel utilization as compared to the standard \mbox{Wi-Fi}, irrespective of~$ \eta $. 
\vspace{-0.55cm}
\subsection{UL and DL traffic with full buffer } 
Figs.~\ref{M1} and \ref{R1} show the performance of victim and non-victim users---in terms of their throughputs and successful channel access percentages with varying $ \eta $---for standard \mbox{Wi-Fi} and proposed scheme, respectively. A similar trend in throughput, as with DL only traffic, can be observed ensuring fairness among the victim and non-victim users for $ \eta \leq \eta_t $, (with $ \eta_t = 0.75 $). For $ \eta > \eta_t $, the above conclusion, of \mbox{LTE-U} OFF period being insufficient for victim user still holds true. Successful channel access also being equal confirms that the throughput fairness was a result of equal access. 
Apart from fairness, for all $ \eta < 1$, a noticeable improvement in the throughput of victim user can be observed. 
\par Fig.~\ref{R2} shows the performance of \mbox{Wi-Fi} AP in terms of DL throughput, DL opportunity, and successful DL transmission by varying \mbox{LTE-U} ON fraction. DL opportunity in case of standard \mbox{Wi-Fi} (also shown in Fig. \ref{M2}) decreases as $ \eta $ increases, since multiple retransmissions to victim user increases the average BO value. While, with the proposed scheme, the \mbox{Wi-Fi} AP cleverly defers from transmitting to the victim user, during \mbox{LTE-U} ON period and avoids unnecessary increase in its average BO value, thereby preventing the decrease in DL opportunity of \mbox{Wi-Fi} AP. DL opportunity is now expected to be at $33\%$ when $ \eta $ is 0  (since all the three nodes get equal chance to access the channel, when \mbox{LTE-U} is OFF), and would increase to $50\%$ when $ \eta $ is~1. A DL opportunity of $ 50\% $ (at $ \eta=1$) signifies that the non-victim user does not dominate the network and \mbox{Wi-Fi} AP still obtain equal chance to transmit as compared to the non-victim user.
\par Successful DL percentage in Fig.~\ref{R2}, which in the standard \mbox{Wi-Fi} was highly impacted by the retransmissions to the victim user, is now---with the proposed scheme---just influenced by the collisions in the network, and hence is very high. Successful DL percentage increases with $ \eta $ until $ \eta_t $. An increase in \mbox{LTE-U} ON time demands for more CFP (to achieve fairness), and higher the CFP duration in a duty cycle, lower the collisions and hence Successful DL percentage increases with $ \eta $.
 For $ \eta > \eta_t $, CFP duration is limited by \mbox{LTE-U} OFF time. Increasing $ \eta $ more than $ \eta_t $, will
 decrease the CFP duration due to a decrease in LTE-U OFF period. This constrained decrease in CFP duration results in a slight decrease in Successful DL percentage (for $ \eta > \eta_t $). 
Further, the DL throughput of the \mbox{Wi-Fi} AP has increased in comparison to the standard \mbox{Wi-Fi}, very distinguishably at higher $ \eta $, which is clearly a result of increase in the DL opportunity and Successful DL opportunity. The decrease in throughput with $ \eta $ is unambiguous---due to the increased \mbox{LTE-U} interference duration.

\par Fig.~\ref{R1_2} shows how the UL throughput varies with the proposed scheme and standard \mbox{Wi-Fi}. Clearly, the very high UL throughput---in standard \mbox{Wi-Fi}---is because the only user not affected by \mbox{LTE-U} is the non-victim user, and is able to send its data too frequently. Now, with the proposed scheme, this perquisite enjoyed by the non-victim user was moderated. An expectation regarding DL throughput would be that it should be less than the UL (half in the above scenario), which is the case at very low duty cycle, but, as the duty cycle increases, this difference decreases. The decrease, in the considered scenario, can be attributed to the fact that CFP allows equal UL and DL opportunities, during the \mbox{LTE-U} OFF period and during the \mbox{LTE-U} ON period, the proposed scheme prevents Wi-Fi AP from degrading its performance and hence allows it to equally compete with the non-victim user for the channel. Increase in $ \eta $, further advocates both the above causes. Nonetheless, the total throughput of the \mbox{Wi-Fi} network with the proposed scheme increased by $ 18\% $, when averaged over $ \eta $ with a maximum value of $ 45.5\% $ (at $ \eta=1 $), compared to standard \mbox{Wi-Fi}, for the considered scenario.

\section{Conclusions and Future Work}
In this paper, we focused on improving the performance of \mbox{Wi-Fi} network in the presence of \mbox{LTE-U} network. We have observed that the performance of \mbox{Wi-Fi}, depending on the interference from \mbox{LTE-U} eNB, can degrade to a large extent. We found that, in cases when \mbox{LTE-U} transmissions partially affect \mbox{Wi-Fi} network, not only the fairness among \mbox{Wi-Fi} users is compromised, but also the \mbox{Wi-Fi} AP's throughput is affected. To address this issue, we proposed a scheme based on PCF, which can be easily adopted by \mbox{Wi-Fi}, to ensure fairness among its users and to improve the performance of \mbox{Wi-Fi} AP. The efficacy of the scheme was proved by sufficient analysis and simulation experiments. As a part of future work, we plan to evaluate the performance of \mbox{multi - LTE-U} and \mbox{multi - Wi-Fi} scenario while employing the proposed scheme. 



\nocite{*}
\bibliographystyle{ieeetr}


\end{document}